\documentclass[aps,reprint,superscriptaddress,
nofootinbib,showpacs,amsmath,amssymb,showkeys,showpacs]{revtex4-1}

\usepackage{latexsym,graphicx,slashed,bm,dcolumn,color}
\usepackage{blkarray}
\usepackage{multirow}
\usepackage{mathtools}
\usepackage{textcomp}
\usepackage{booktabs}
\usepackage{array,mathtools,amssymb}

\usepackage{bm}
\usepackage{enumitem}
\usepackage{etoolbox}

\makeatletter
\patchcmd\frontmatter@PACS@format{\addvspace{11\p@}}{}{}{}
\pretocmd\frontmatter@keys@format{\addvspace{11\p@}}{}{}
\patchcmd{\titleblock@produce}
  {\@pacs@produce\@pacs\@keywords@produce\@keywords}
  {\@keywords@produce\@keywords\@pacs@produce\@pacs}
  {}{}
\makeatother

\begin{document}

\title{Monte Carlo simulation for Ultra-cold neutron experiments searching for neutron - mirror neutron oscillation}

\author{Riccardo Biondi}
\email{riccardo.biondi@aquila.infn.it}
\affiliation{Dipartimento di Fisica e Chimica, Universit\`a di L'Aquila, 67100 Coppito, L'Aquila, Italy} 
\affiliation{INFN, Laboratori Nazionali del Gran Sasso, 67010 Assergi,  L'Aquila, Italy}


\begin{abstract}
Neutron oscillation into mirror neutron, a sterile state exactly degenerate in
mass with the neutron, could be a very rapid process, even faster than the neutron decay itself.
It can be observed by comparing the neutron lose rates in an
ultra-cold neutron trapping experiment for different experimental magnetic fields. 
We developed a Monte Carlo code that simulates many of the features of this kind of experiment with non-uniform magnetic fields.
The aim of the simulation is to provide all necessary tools, needed for analyzing experimental results for neutron traps with different geometry
and different configurations of magnetic field.
This work contain technical details on the Monte Carlo simulation used for the analysis in \cite{Nostro} not presented in it.
\end{abstract}



\maketitle

\section{Introduction}
The concept of mirror world \cite{LY1,LY2} suggested long time ago for parity restoration attracted a considerable interest during last times 
(for reviews see refs. \cite{IJMPA1,IJMPA2,IJMPA3,IJMPA4}).
If mirror world exits, then all ordinary particles: the electron $e$, proton $p$, neutron $n$, photon $\gamma$, 
neutrinos $\nu$ etc. must have invisible twin degenerate in mass: $e'$, $p'$, $n'$, $\gamma'$, $\nu'$ etc. 
which have no strong and electroweak interactions but, their own analogous interaction.
It means that if ordinary particles are described by  the Standard Model $G = SU(3)\times SU(2)\times U(1)$ or its extension, 
then mirror particle will be described by another copy of standard model $G' = SU(3)'\times SU(2)'\times U(1)'$ or its respective extension.
Mirror matter interacting with ordinary matter via gravity provides an interesting and testable candidate for dark matter 
\cite{BDM1,BDM2,BCV1,BCV2,BCV3,BCV4}.
However, particles of the two sectors can have some common interactions other then gravity \cite{PLB981,PLB982,PLB983,PLB984,PLB985}.
Big Bang nucleosynthesis bounds and cosmological consistency on mirror dark matter require that
these interactions should be rather feeble in order not
to bring the two sector in thermal equilibrium in early universe \cite{BCV1,BCV2,BCV3,BCV4}. 
Nevertheless they can be important for dark matter direct detection \cite{DAMA1,DAMA2,DAMA3,DAMA4}.
In addition, they can induce mixing phenomena between neutral ordinary and mirror particle, which could be testable.

For example, the three ordinary neutrinos $\nu_{e,\mu\,\tau}$ can oscillate into their mirror partners $\nu'_{e,\mu\,\tau}$  which are
in fact natural candidates for sterile neutrinos 
\cite{M-neutrinos1,M-neutrinos2,M-neutrinos3,M-neutrinos4}. 
This mixing can emerge via effective interactions which violate $B-L$ symmetries in both ordinary
and mirror sectors.  
In the early Universe, these interactions would induce CP violating processes between ordinary and mirror particles
which can co-generate baryon asymmetries in both sectors \cite{BB-PRL1,BB-PRL2,BB-PRL3,BB-PRL4}.  
In this way, one could naturally explain the relation $\Omega'_B /\Omega_B \simeq 5$    
between the baryon and dark matter fraction in the Universe \cite{BB-PRL1,BB-PRL2,BB-PRL3,BB-PRL4,IJMPA1,IJMPA2,IJMPA3,IJMPA4}.

As it was shown in refs. \cite{BB-nn',More}, the oscillation between neutron $n$ and its mirror twin $n'$, emerging 
due to the mass mixing term: $\epsilon (\bar{n} n' + \bar{n}' n)$, can be a fast process and it can be even faster than neutron decay.
Energy conservation does not allow  $n \to n'$ transition in stable nuclei, and so, no limit emerges on the oscillation time from
the nuclear stability \cite{BB-nn'}. 
For free neutrons $n$--$n'$ oscillation is influenced by matter and magnetic fields \cite{BB-nn',More}. 
By that reason,
existing experimental limits and astrophysical bounds still allow oscillation time $\tau = \epsilon^{-1}$ as small as few seconds 
\cite{BB-nn'}, in drastic difference from the case of neutron-antineutron oscillation for which experimental lower bounds correspond
to few years \cite{Phillips1,Phillips2,Phillips3}.
It is interesting that $n\to n'$ transitions faster than the neutron decay can have intriguing implications for the origin and
propagation of ultra-high energy cosmic rays \cite{UHECR1, UHECR2}. 
In laboratory conditions, $n-n'$ oscillations can be tested in experiments searching for neutron disappearance $n\to n'$ and 
regeneration $n\to n' \to n$ \cite{BB-nn',Pokot,ORNL1,ORNL2} as well as via non-linear effects on the neutron spin precession \cite{More}.

The ultra-cold neutron (UCN) storage experiments have good capabilities for testing the neutron disappearance phenomena. 
The status of presently existing UCN sources for fundamental physics measurements were recently reviewed in ref. \cite{Bison}.
In UCN traps the $n \to n'$ oscillations can show up by measuring the magnetic field dependence of the UCN losses. 
In the last years, many UCN storage experiments were dedicated to search for $n-n'$ oscillations 
\cite{Ban,Serebrov,Bodek,Serebrov-New,Altarev,Nostro}.
The experiments \cite{Ban,Serebrov,Bodek,Serebrov-New}, compared the UCN loss rates 
in \emph{zero} (i.e. small enough) and {\it non-zero} (large enough) experimental magnetic field
assuming that the Earth has no mirror magnetic field \cite{BB-nn'}.
They obtained lower bounds on the oscillation time, with the strongest bound  $\tau > 414$~s  at 90\% CL of ref. \cite{Serebrov} 
adopted by the Particle Data Group \cite{PDG}.

These limits are not valid if there exists a non-vanishing mirror magnetic field $B'$ at the Earth \cite{More}. 
It would suppress $n-n'$ oscillation even if the ordinary magnetic field is screened ($B = 0$). 
On the other hand, if experimental magnetic field is tuned as $B\approx B'$, then $n-n'$ oscillation would be resonantly enhanced \cite{More}. 
A detailed analysis of experimental data from \cite{Serebrov-New} indicates towards more that $5\sigma$ deviation from the
null hypothesis \cite{Berezhiani:2012rq} which can be interpreted as a signal for $n-n'$ oscillation with $\tau \sim 10$ s 
in the presence of mirror 
magnetic field $B'\sim 0.1$ G.  
The mirror magnetic field at the Earth can emerge if the Earth captures some amount of mirror matter.
Then, Earth rotation would drag mirror electrons inducing circular electric currents sourcing the mirror magnetic fields, 
which can be further amplified by dynamo mechanism \cite{More,BDT}.
The experiment \cite{Altarev} tested $n-n'$ oscillation in the presence of mirror magnetic fields.
Its results, gives the lower limit $\tau > 12$~s for any $B'$ less than 0.13 G.
Somewhat stronger limits on $n-n'$ oscillation parameters have been recently reported in \cite{Nostro}.
These limits \cite{Altarev,Nostro} restrict the parameter space which 
can be responsible for the above $5\sigma$ anomaly but do not cover it completely. 
Therefore new more precise experiment are needed to verify this anomaly.


To maximize the chances of finding a signal of $n-n'$ oscillation in a UCN trap experiment, we have to apply the 
magnetic field as close as possible to the unknown mirror magnetic field.
The most precise way would be to apply uniform magnetic field (as in the case of  refs. \cite{Ban,Serebrov,Bodek,Altarev}
and horizontal measurements of ref. \cite{Serebrov-New}) and to vary its value with very small steps for capturing the resonance.
But such a procedure would require quite a long overall time of measurements.
On the other hand, one could apply a non-uniform magnetic field (as in case of measurements with vertical magnetic field refs. 
\cite{Serebrov-New,Nostro}).
In this case, if the mirror field value falls  within the spread of the applied magnetic field profile inside the trap, 
the neutrons moving in the trap with some probabilities cross this resonant areas where they oscillate very efficiently.
This can give us some good possibility to catch the effect of the resonant amplification and give some gain in experimental time and 
statistics, since there will be no need to scan the magnetic field with very small steps.

This needs precise evaluation of oscillation probability for neutrons moving in non-uniform magnetic field.
For this purpose we developed a Monte Carlo code and used it 
for the analysis of the  results of the experiment \cite{Nostro}.
This tool was applied also to reinterpret the implications for parameter space from ref. \cite{Berezhiani:2012rq} 
based on the analysis of the experiment \cite{Serebrov-New}, 
since  in \cite{Berezhiani:2012rq} the non-uniformity of the magnetic field was not correctly taken into account and
a specific empirical formula was used for averaging oscillation probability.
As it was shown in \cite{Nostro}, the non-homogeneity of the magnetic field
substantially altered the relevant space for oscillation parameters.

In this paper we report the main features of the Monte Carlo code.  
It was written with the purpose of creating a tool that can be used in the analysis of any $n-n'$ oscillation in a UCN trap experiment with 
non-uniform magnetic field and can be applied for material or magnetic UCN traps of different geometrical configurations.
The paper is organized as follows: first we discuss $n-n'$ oscillation in the presence 
of mirror magnetic field and the experimental setup. Then we describe key features of the simulation such as the UCN diffusion, wall scattering,
loss probability and the behavior of $n-n'$ oscillation probability.
Finally we confront our results with the previous calculations and draw our conclusions.

\section{Neutron - Mirror Neutron Oscillations in Homogeneous Magnetic Field}
Data form UCN experiment can be tested for magnetic field dependence of UCN losses, as a probe for $n-n'$ oscillation. 
In fact, if free neutrons propagate through vacuum in a region where ordinary $\mathbf{B}$ and mirror $\mathbf{B}'$ 
magnetic fields are both non-zero and have arbitrary orientation, 
the following Schr\"odinger equation with a $4\times 4$ Hamiltonian, describes 
$n-n'$ oscillation phenomenology 
\begin{equation}
\frac{d\Psi}{dt} = H_{nn'} \Psi \;\;\;\; with \;\;\;\;
 H_{nn'} = \left( \begin{array}{cc}
               \mu \mathbf{B}  \cdot \mathbf{\sigma} & \epsilon \\
               \epsilon  & \mu \mathbf{B}' \cdot \mathbf{\sigma}
              \end{array} \right) \label{eq:schr}
\end{equation}
where $\Psi= (\psi_n(t) , \psi_{n'}(t))$, is the wave function of $n$ and $n'$ in two spin states, 
and $\mathbf{\sigma}=(\sigma_x,\sigma_y,\sigma_z)$ are the Pauli matrices.  
Exact derivation of $n \to n'$ transition probability is given in ref. \cite{More}.  
Following \cite{Berezhiani:2012rq}, in homogeneous fields  $\mathbf{B}$ and $\mathbf{B}'$, the probability of $n-n'$ oscillation after  
neutron flight time $t$ can be reduced to: 

%
%

\begin{align}
 P_{\mathbf{B} \mathbf{B}' }(t)   = & \frac{ \sin^2[(\omega-\omega')t] }{2  \tau^2 (\omega -  \omega^\prime)^2 } \left( 1 + \cos\beta \right) + \nonumber \\
 & \frac{ \sin^2[(\omega+\omega')t] }{2  \tau^2 (\omega +  \omega^\prime)^2 } \left( 1 - \cos\beta \right)
 ,  \label{eq:P}
\end{align}
where $\beta$ is the angle between $\mathbf{B}$ and $\mathbf{B}'$, and 
$\mu = -6 \cdot 10^{-12}\,$eV/G being the neutron magnetic moment\footnote{we use natural units $\hbar = c =1$}, and
$\tau = \epsilon^{-1}$, $ \omega = \frac12 | \mu B|$ and  $ \omega' = \frac12 | \mu B' |$.
Given $B=\vert \mathbf{B} \vert$ and $B'=\vert \mathbf{B}' \vert$ the oscillation probability  
eq. (\ref{eq:P}) becomes maximal or minimal respectively when $\mathbf{B}$ and $\mathbf{B}'$ are parallel or anti-parallel,  
$\cos\beta = \pm 1$: 


\begin{equation}
 P_{BB' }^{(\pm)}(t) = \frac{\sin^2[(\omega \mp \omega')t]}{\tau^2 (\omega \mp \omega')^2} ,  \label{maxmin}
\end{equation}

Experimental magnetic field can be controlled and thus the dependence of $n-n'$ conversion probability on $\mathbf{B}$ can be tested.
We can reverse  magnetic field direction $\mathbf{B} \to - \mathbf{B}$ (i.e. $\beta \to \pi - \beta$)
or measure counts with zero magnetic field. Therefore we can define:

\begin{align}
 D_{BB'} \cos\beta & \equiv \frac{P_{\mathbf{B}\mathbf{B}'} - P_{-\mathbf{B}\mathbf{B}'} }{2} \nonumber \\
 \mathcal{P}_{BB'} &  \equiv P_{\mathbf{B}\mathbf{B}'} + P_{-\mathbf{B}\mathbf{B}'}  \nonumber \\
 \mathcal{P}_{0B'} & \equiv P_{\mathbf{0}\mathbf{B}'}. \label{Pdef}
\end{align}
$D_{BB'}$ depend on $\beta$ and is non-zero only if $\cos\beta \neq 0$, while $\mathcal{P}_{BB'}$ and $\mathcal{P}_{0B'}$
are independent from the orientation of $\mathbf{B}$ and $\mathbf{B}'$.

%
For trapped UCN, oscillation probability eq. (\ref{eq:P}) should be averaged over the distribution of neutron flight times 
$t$ between wall collisions. 
For averaging the sinusoidal factors in eq. (\ref{maxmin}) we can use the empirical formula from ref. \cite{Berezhiani:2012rq}: 
\begin{equation}
 \langle \sin^2(\omega t) \rangle_t = S(\omega)  = 
\frac12 \left[1 - e^{-2\omega^2 \sigma_f^2} \cos(2 \omega t_f) \right], \label{eq:empirical}
\end{equation}
where $t_f = \langle t \rangle$ is the neutron mean free-flight time between wall collisions 
and $ \sigma_f^2  = \langle t^2 \rangle - t_f^2$. is its variance. 
This parameters have to be computed via Monte Carlo simulation for each trap, depending on its geometry and size.
Eq. (\ref{eq:empirical}) gives a correct asymptotic behavior for the average probabilities eq. (\ref{maxmin}): 
\begin{equation}
 \bar{P}_{BB' }^{(\pm)} = \frac{S(\omega \mp \omega')}{\tau^2 (\omega \mp \omega')^2}. 
\end{equation}
Calculation of $\bar{P}_{BB' }$ with the analytic approximation 
eq. (\ref{eq:empirical}) agree at the level of percent to the one obtained via Monte-Carlo simulations if we consider a trap
with homogeneous magnetic field.

As stated before, this oscillation phenomenon can be tested via magnetic field dependence of UCN losses.
The regular UCN losses during the storage:  $\beta$-decay, 
wall absorption or upscattering are magnetic field independent, so
the number of neutrons $N(t_\ast)$ survived after being stored in the trap for a certain time $t_\ast$
should not depend on $\mathbf{B}$, in the standard physics paradigm.

Nevertheless, if a neutron, between two consecutive wall collisions, oscillates into a mirror neutron state $n'$, 
then per each collision it has some non-zero probability of escaping from the trap. 
Thus, the amount of survived neutrons in the UCN trap with applied magnetic field $\mathbf{B}$ after a time $t_\ast$ is given by 
$N_{\mathbf{B}}(t_\ast) =  N(t_\ast)  \exp( -n_\ast  \bar{P}_{\mathbf{B}\mathbf{B}'})$,
where $\bar{P}_{\mathbf{B}\mathbf{B}'}$ is the average probability of $n-n'$ conversion between the wall scatterings 
and $n_\ast = n(t_\ast)$ is  the mean number of wall scatterings.
If we reverse the magnetic field  direction, $\mathbf{B} \to -\mathbf{B}$, then we have 
$N_{-\mathbf{B}}(t_\ast) = N(t_\ast)  \exp (- n_\ast  \bar{P}_{-\mathbf{B}\mathbf{B}'}) $. 
Since  the common factor $N(t_\ast)$ cancels in the  neutron count ratios,  
asymmetry between $N_{\mathbf{B}}(t_\ast)$ and $N_{-\mathbf{B}}(t_\ast)$ is:
\begin{equation}
 A_{\mathbf{B}}(t_\ast) = \frac{N_{-\mathbf{B}}(t_\ast) -
  N_{\mathbf{B}}(t_\ast)} {N_{-\mathbf{B}}(t_\ast)+N_{\mathbf{B}}(t_\ast)} = n_\ast \bar{D}_{BB'}  \cos\!\beta \, , \label{eq:AB}
\end{equation}
assuming $n_\ast \bar{D}_{\mathbf{B}\mathbf{B}'} \ll 1$. 
However, one can compare $N_B(t_\ast) = \frac12 \big[N_{\mathbf{B}}(t_\ast) + N_{-\mathbf{B}}(t_\ast)\big]$ 
with the counts $N_0(t_\ast)$ measured under zero experimental magnetic field: 
\begin{equation}
 E_{\mathbf{B}}(t_\ast) = \frac{N_0(t_\ast) - N_{\mathbf{B}}(t_\ast)}
 {N_{0}(t_\ast)+N_{\mathbf{B}}(t_\ast)} = n_\ast (\bar{\mathcal{P}}_{BB'} - \bar{\mathcal{P}}_{0B'})  , \label{eq:EB}
\end{equation}
$E_{\mathbf{B}}$ traces directly the difference between the probabilities in zero and non-zero magnetic fields. 
which depend only on its modulus $B = \vert \mathbf{B} \vert$, and it should be resonantly amplified if $B\approx B'$. 
Since the parameter $n_\ast$ indicate the averaged number of wall collision for each UCN, its estimation it is very
important for setting limits on the oscillation probability.

Measuring $E_B$ for many values of $B$, one can obtain limits 
on $n-n'$ oscillation time $\tau$. On the other hand, from $A_{\mathbf{B}}$ one in fact 
measures the value $\tau_\beta = \tau/\sqrt{\vert \cos\beta \vert}$, i.e. 
oscillation time corrected by the unknown 
angle $\beta$ between ordinary and mirror magnetic fields  $\mathbf{B}$ and $\mathbf{B}'$.
In ideal conditions,
the effects of regular neutron loses cancel out from the ratios $A_{\mathbf{B}}$ and $E_B$, and measuring them for 
different values of  $\mathbf{B}$, it is possible to obtain 
pretty robust limits on $\tau$ and $\tau_\beta$ as a function of mirror magnetic field $B'$.

\section{Experimental Scheme}

\begin{figure}[!ht]
 \begin{center}
\includegraphics[width=1.0\columnwidth]{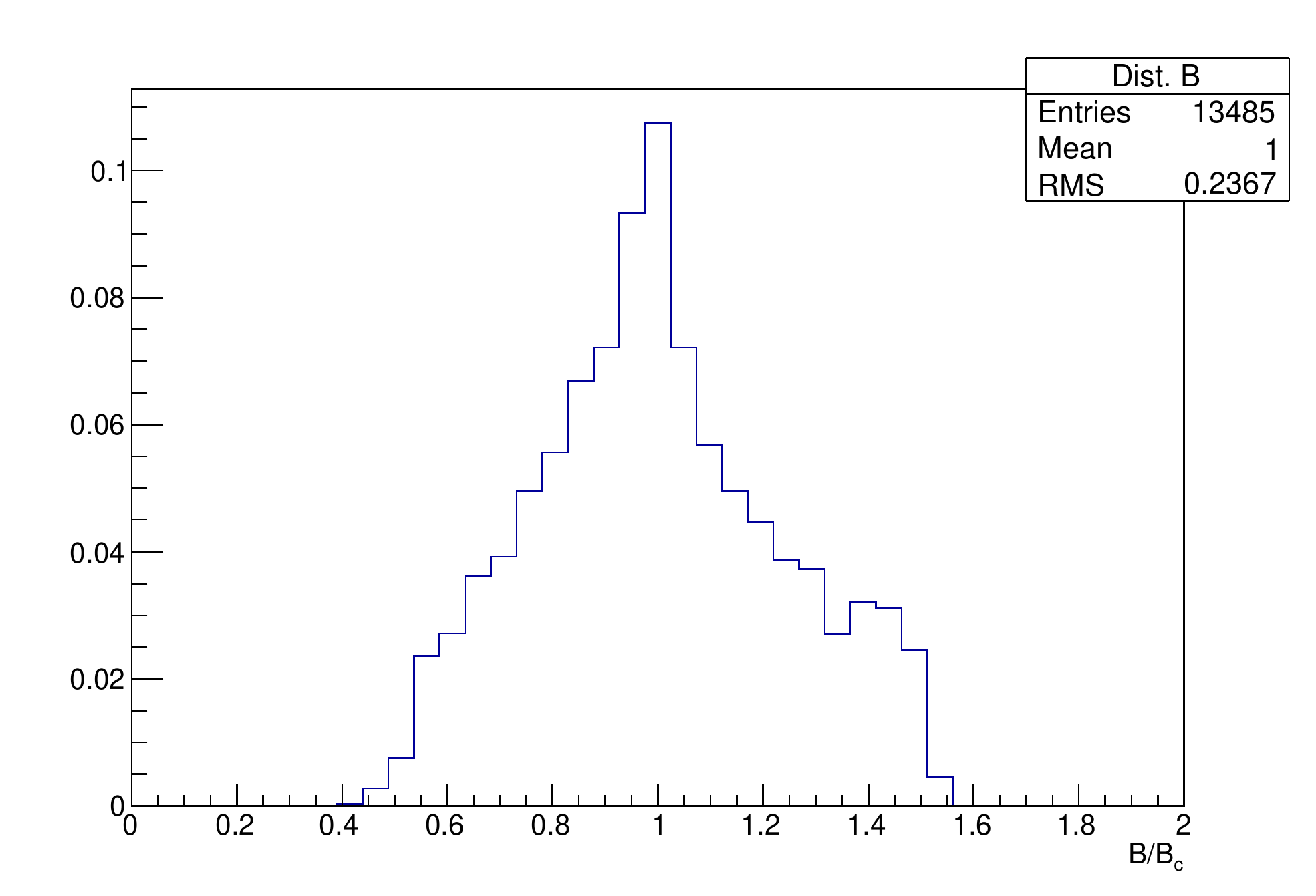}
\includegraphics[width=1.0\columnwidth]{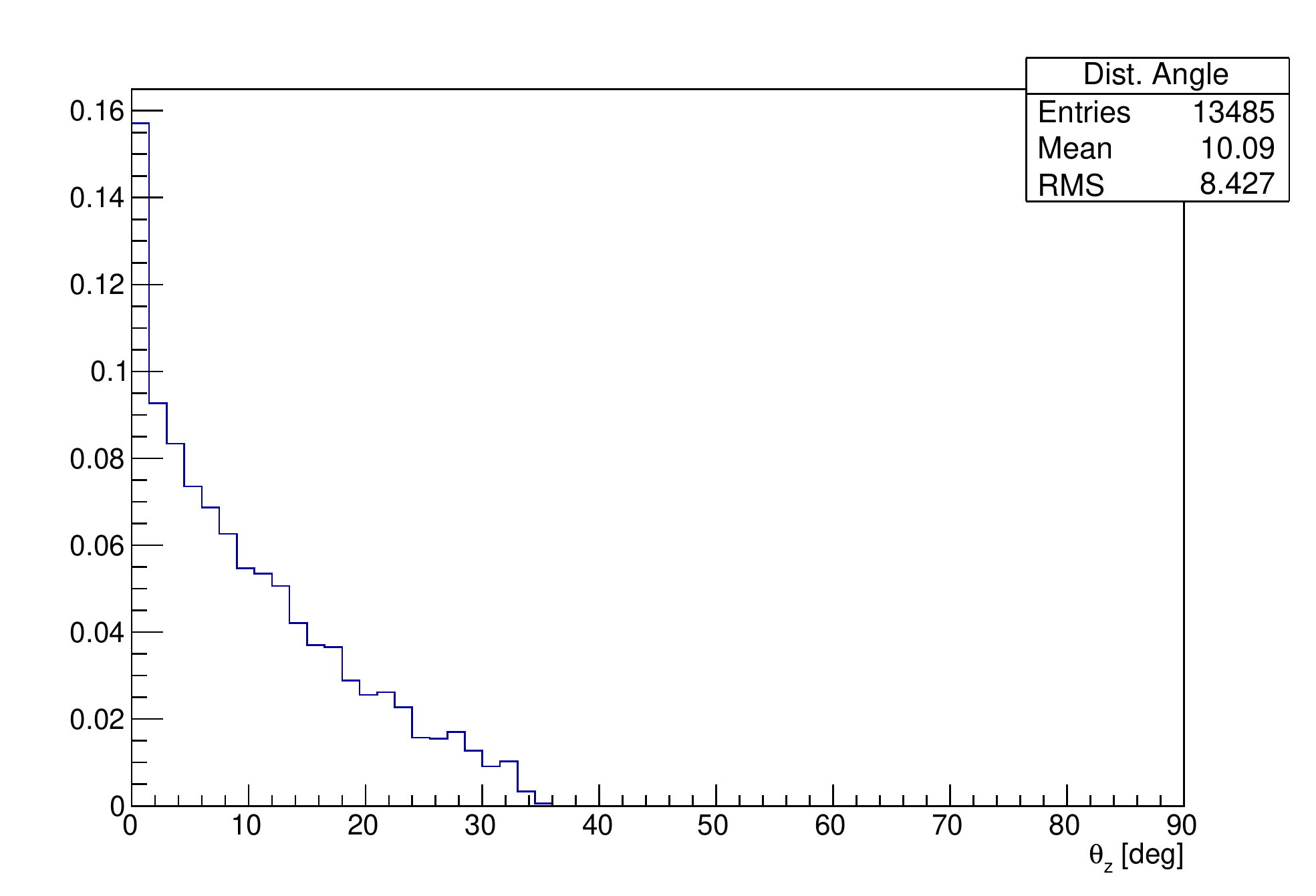}
\caption{\label{fig:dist} 
Magnetic field in the trap used in experiment \cite{Nostro}.
{\it Upper panel:} distribution of the absolute value of the magnetic field around the central value $B_c$.   
{\it Lower panel:} distribution of the deviation angles of vector $\vec{B}$ relative to $z$ axis. }
\end{center}
\end{figure}

\begin{figure}[ht]
  \centerline{\includegraphics[width=0.85\columnwidth]{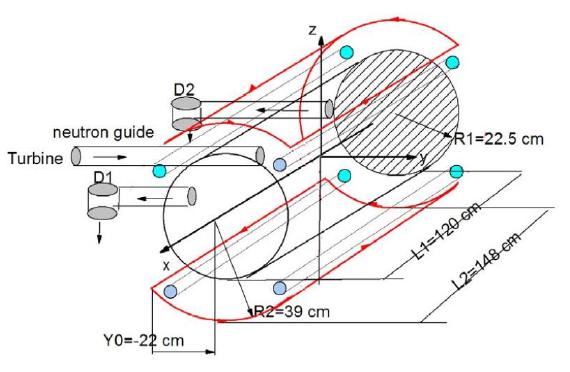}}
 \caption{
 Scheme of the UCN trap used in \cite{Nostro},the red lines, indicate the position of the electric 
circuits that generate vertical magnetic field inside the trap.}
  \label{fig:5_trap}
\end{figure}

UCN experiment such as \cite{Serebrov,Serebrov-New,Nostro} employed a trap of $190\,\ell$ volume capable of storing about 
half million neutrons, located inside a shield screening the Earth magnetic field. 
A controlled magnetic field is then induced by a system of solenoids.
It can be oriented in both horizontal and vertical direction, but, in the vertical field setup it is not possible to have uniform magnetic
field inside the trap. 
In this case, the induced magnetic field $\vec{B}$ has vertical direction practically everywhere inside the trap,
but its magnitude $B= \vert \vec{B} \vert$ 
is in-homogeneously distributed, varying from the value $B_c$ in the geometrical center by about $\pm 0.5\, B_c$ at peripheral regions. 
The distribution of magnetic field inside the trap used in the experiment \cite{Nostro} is shown in fig. \ref{fig:dist}.  
The scheme of the installation used in the experiments \cite{Nostro} is shown in Fig. \ref{fig:5_trap}.
The trap used in \cite{Serebrov,Serebrov-New,Nostro} is a horizontal cylinder with diameter 45 cm and length 120 cm. 
It was made of copper with the inner surface coated by beryllium.
The critical velocity of this coating is 6.8 m/s  which corresponds to the Beryllium optical potential: 2.7 $\times$ $10^{-7}$ eV.
The magnetic shielding of the installation consists of four layers of permalloy, and the
residual magnetic field inside the shielding is about 2 nT, which is low enough to carry out the search for $n-n'$ transitions. 

Every measurement, lasting about 10 minutes, consists of five phases: 
monitoring, filling, storage, emptying and background.
In the monitor phase, the entrance valve is open and neutrons flow into the trap via the UCN guide while the exit valves 
that communicate with the two detectors $D_1$ and $D_2$ (fig. \ref{fig:5_trap}) remain open. 
Their counts during the monitor phase are used as estimators of the incident UCN flux and for checking its stability.
Then the exit valves are closed for the filling phase, 
after which the entrance valve is closed and neutrons are kept in the trap for a storage time $t_{s}$.  
Then the exit valves are again opened and the survived UCN are counted 
by two detectors during the emptying phase.
The final background phase checks that no excess of neutrons remains in the trap
which could influence the following measurement.

\section{UCN Motion}
\begin{figure}[!ht]
 \begin{center}
 \includegraphics[width=0.85\columnwidth]{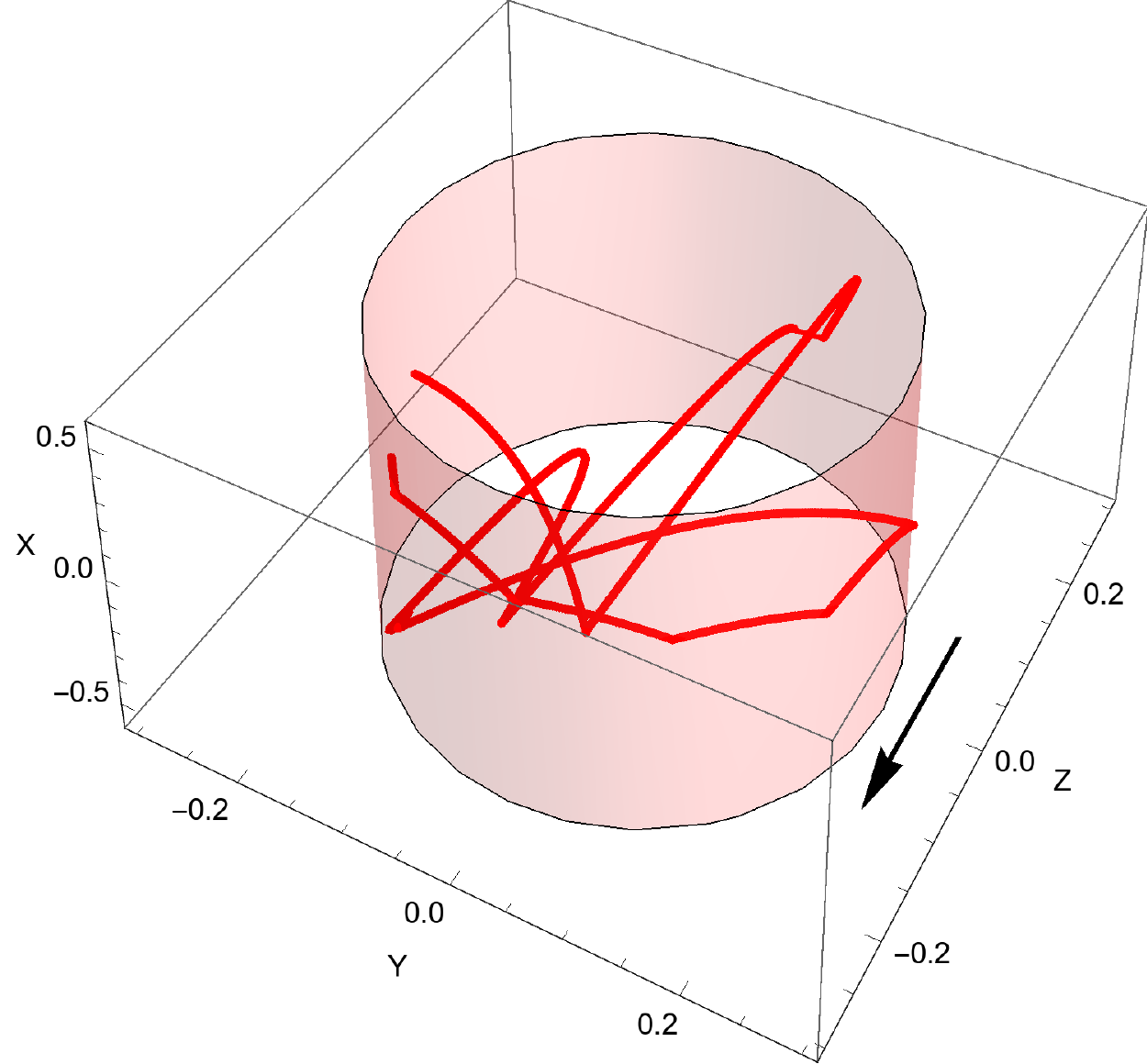}
\caption{\label{fig:5_UCN_path} UCN path inside the trap. The black arrow indicates the orientation of gravity acceleration. }
\end{center}
\end{figure}

\begin{figure}[!ht]
 \begin{center} 
 \includegraphics[width=0.49\columnwidth]{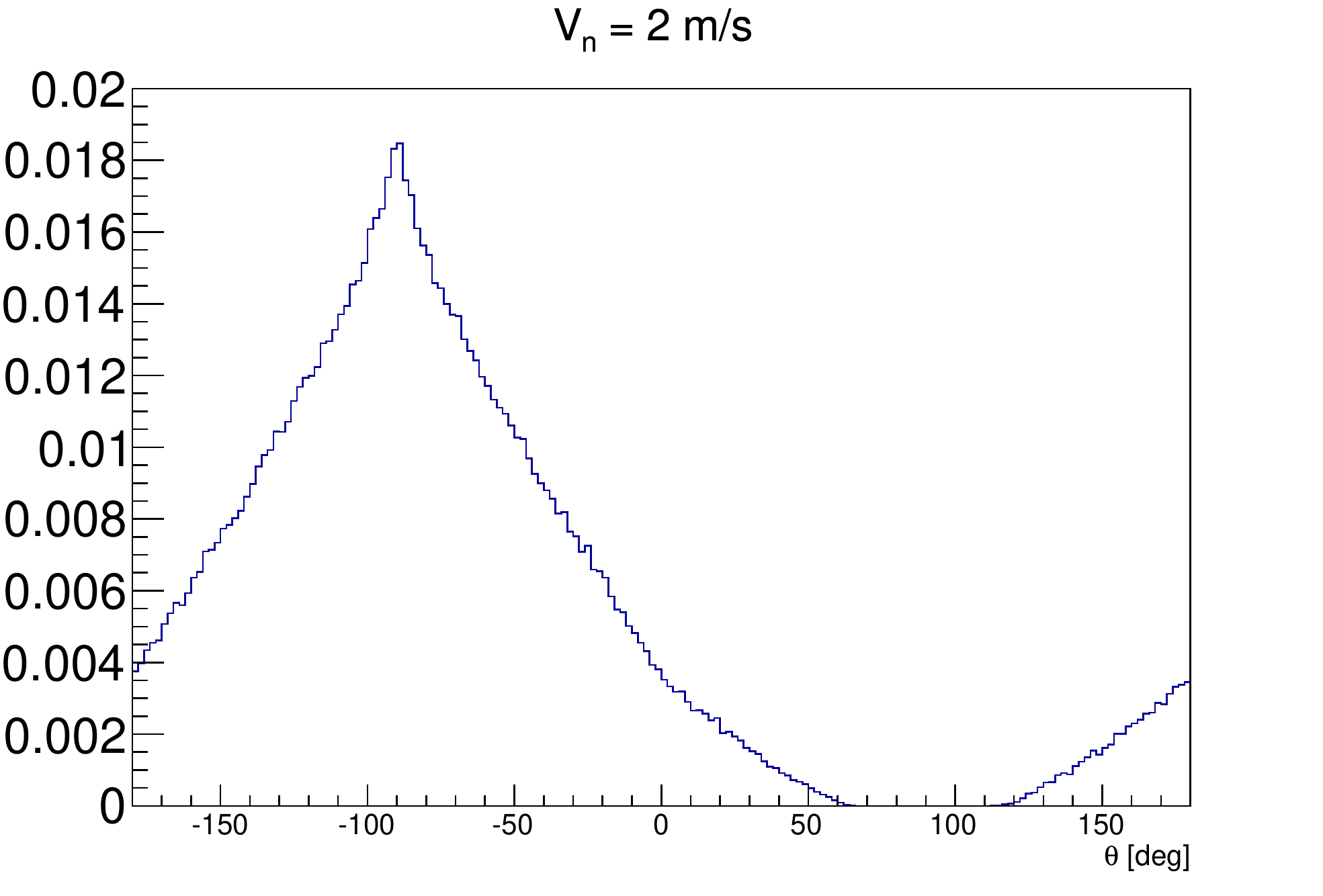} 
 \includegraphics[width=0.49\columnwidth]{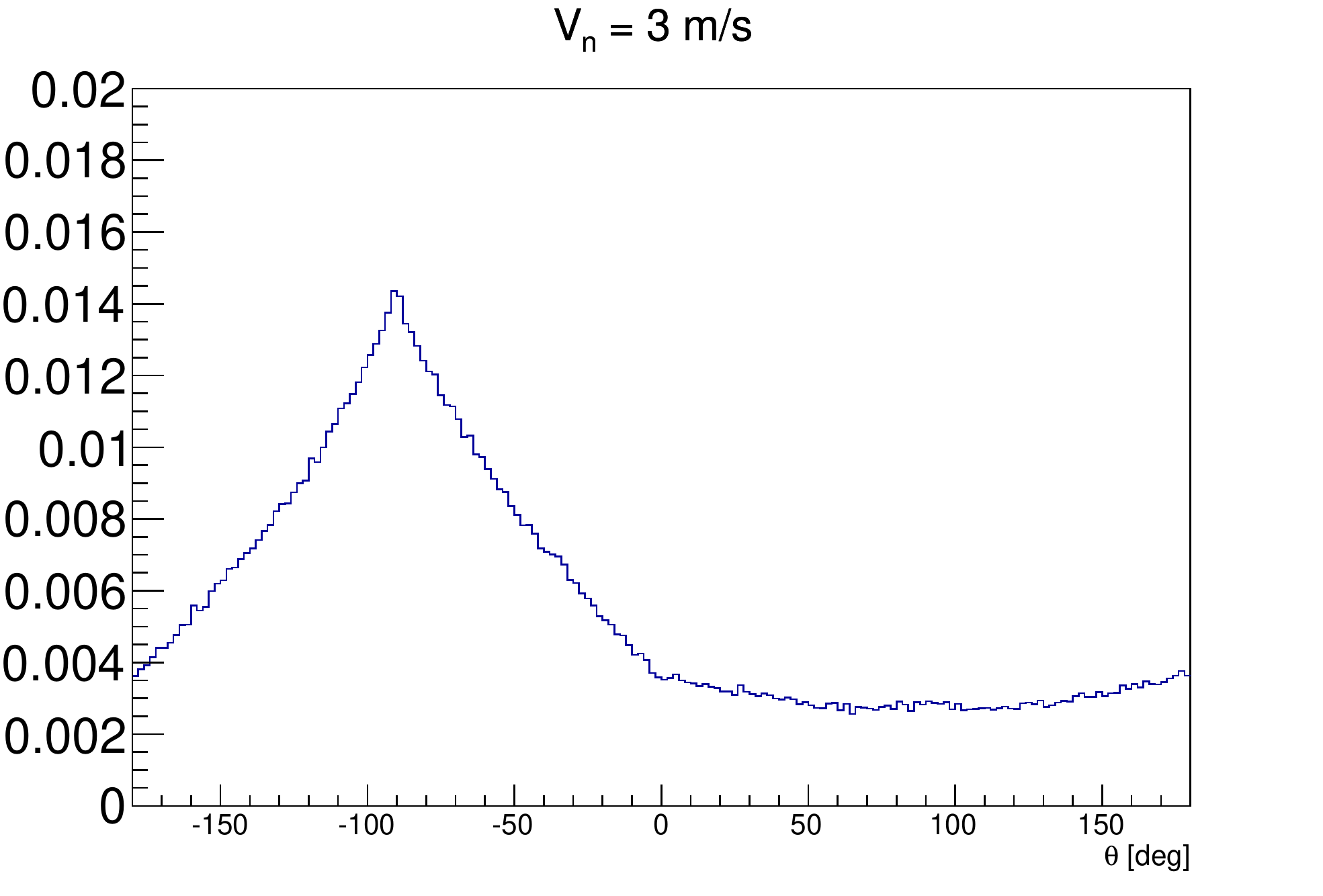}
 \includegraphics[width=0.49\columnwidth]{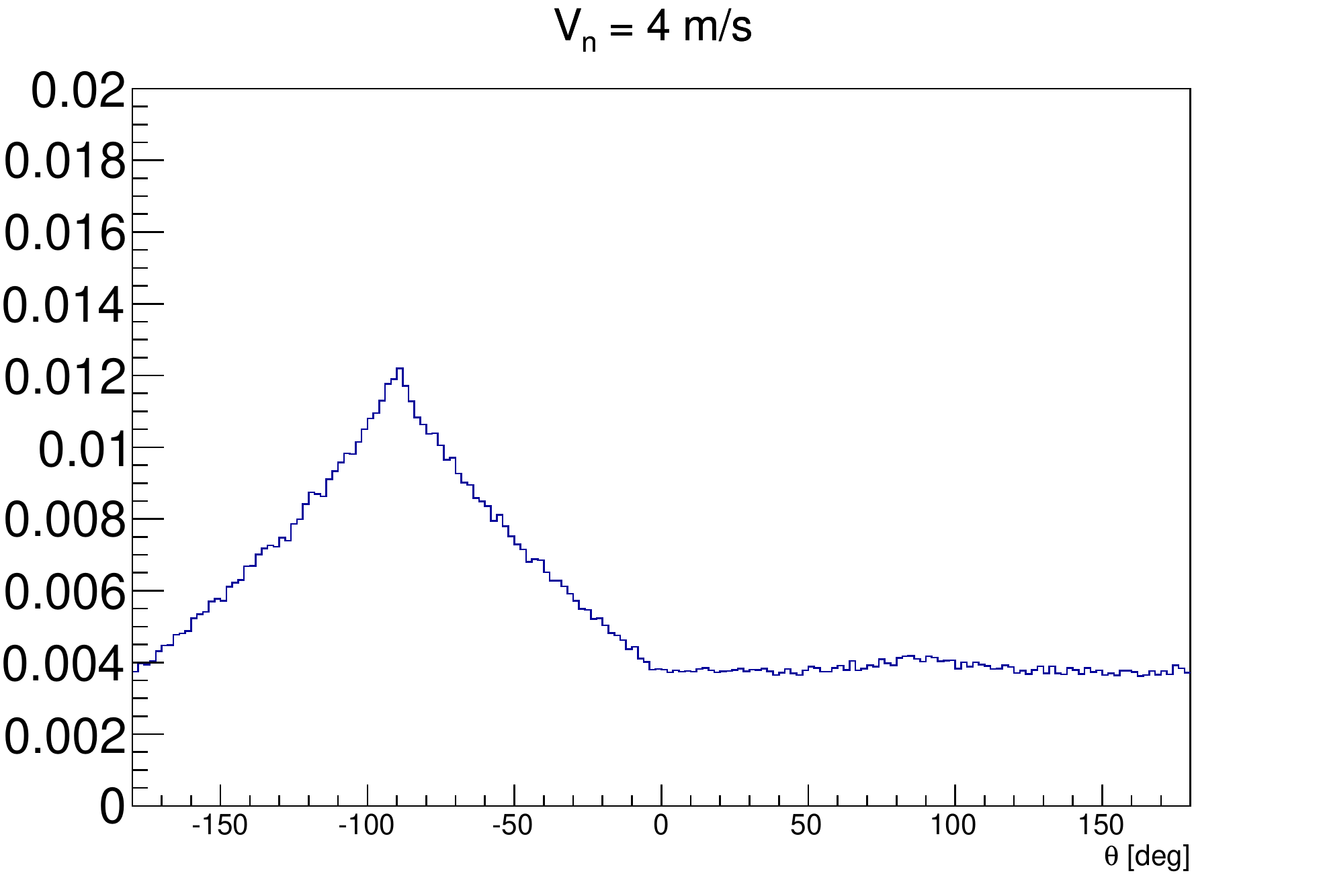} 
 \includegraphics[width=0.49\columnwidth]{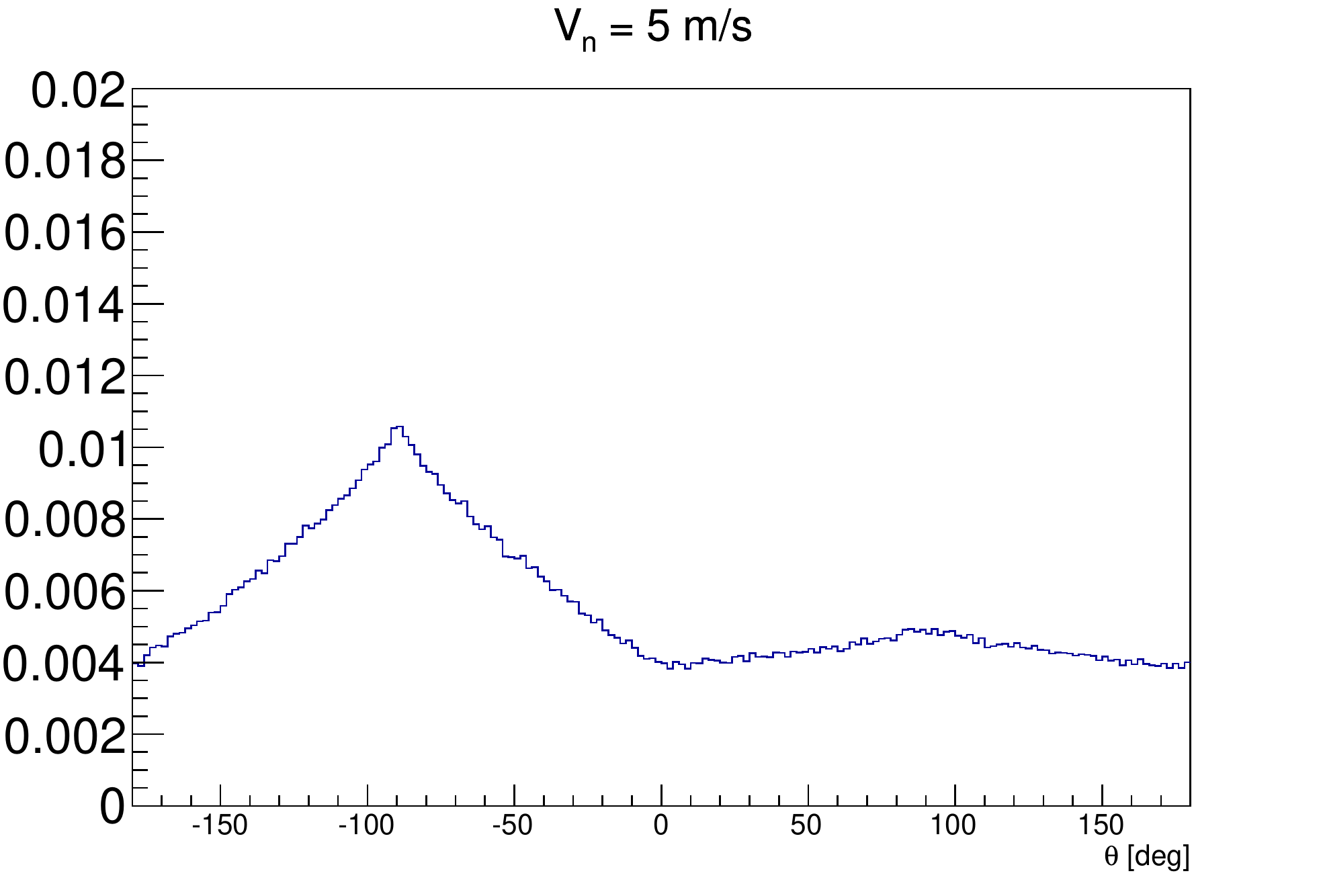}
\caption{\label{fig:coverage} Coverage of UCN hits with the cylindric surface of the trap. 
Normalized distribution in terms of angles with the positive direction of the Y axis, for different velocity classes measured with $t_S$ = 100 s.}
\end{center}
\end{figure}

The program is written in C++ and implemented with ROOT, its core, consists in a simple numerical
integration with the trapezoidal method of the motion of a single UCN that enters inside the trap from the UCN entrance 
(see fig. \ref{fig:5_trap}) with a randomly distributed velocity according to the UCN spectra from: \cite{UCN_spectra}. 
Then it moves inside the trap under the effect of gravity acceleration for a certain storage time $t_S$. 
A key feature is the treatment of the collisions between UCN and the trap.
The modification of the direction of motion induced by the elastic collision, combined with the discretization of time, 
induced an approximation error that grows with $t_S$ and manifest itself with a progressive loss of energy of the UCN. 
To avoid this, for each collision the squared module of the UCN neutron was computed  imposing energy conservation:
\begin{equation}
 v^2  = 2 \left( E_0  - g h \right) 
\end{equation}
where $E_0$ is the initial energy of the UCN normalized over the its mass, $v$ is its velocity, h is the 
height measured from the bottom of the trap and $g$ is gravity acceleration.
With this trick we can have a good approximation even for $t_S > 500$ s  with an energy loss less then 1 \%.
Regarding UCN-trap scattering, half of them (randomly chosen) has been treated as specular scattering and
the other half, as diffuse scattering \cite{UCN_reflection}.
In fig. \ref{fig:5_UCN_path} is shown a typical path of a simulated UCN.
The escape probabilities $P_{esc}$ for each wall collision depends on the orthogonal component of the UCN 
velocity $v_\bot$ \cite{UCN_reflection,Golub} reads:
\begin{equation}
 P_{esc}(v_\bot) =  2 \, \eta  \, \frac{|v_\bot|}{\sqrt{v_{max}^2 - v_\bot^2 } + 2 \eta  \, |v_\bot| }  \label{eq:escape}
\end{equation}
where $v_{max} = 6.8$ m/s and $\eta$ are a parameters that depend on the material and the geometry of the trap, for the trap used in:
 \cite{Serebrov,Serebrov-New,Nostro} we have $\eta \simeq 2.2 \times 10^{-4}$.
The $2 \eta  \, |v_\bot|$ term in the denominator has been added to avoid the probability to diverge when $v_\bot$ $\sim$ $v_{max}$.
In the simulation we also take into account the $\beta$-decay, 
to this aim, in every numerical step each neutron has a decay probability given by $P_\beta$ = $dt/\tau_n$,where dt is 
the temporal step of the numerical integration and $\tau_n = (880.2 \pm 1.0)$ s \cite{PDG} is the free neutron mean lifetime.
If a neutron decays or escapes during it's motion inside the trap, it is discarded and not taken into account for the calculation of parameters 
of interest, because UCN lost for regular reasons are not relevant for the count asymmetry eq. (\ref{eq:AB}) and (\ref{eq:EB}).
It is possible to study the coverage of the hits of UCN with the surface of the trap for different velocity classes. 
Results are shown in fig. \ref{fig:coverage}, as it is easy to guess the coverage is not symmetric due to the effect of gravity, 
so especially for lower velocity class the coverage is higher  in the bottom part\footnote{from -180\textdegree $\;$ to 0\textdegree } of the trap. 
Coverage is important because, the value of magnetic field at the surface of the trap is the one the matter for $n\to n'$ transition.  
With an in-homogeneous magnetic field, we have different values of $B$ for different regions, so, hitting the trap in the areas where the
magnetic field have a value closer to the resonance will convert neutron more efficiently.
For the calculation of $n_\ast$, the code has been extended to simulate the three main phases of any measurements: 
filling, storage and emptying:
\begin{itemize}
 \item \textbf{Filling:} In this phase, UCN are born at the entrance referring to fig. \ref{fig:5_trap} and they move inside the trap
       for a randomly chosen time between 0 and $t_F$ with flat distribution, which is the experimental duration of the filling phase. 
       If during this time interval the neutron decays, escapes or hit the entrance valve, it is considered as lost and not taken into account 
       for the calculation of $n_\ast$. The entrance valve is a circular surface of 8 cm radius. 
 \item \textbf{Storage:} UCN that survived the filling phase enter in the storage one. Here all  entrance valves are closed, 
       so they move inside the trap for a time $t_S$ and, as in the previous phase, if they decay or absorbed at the wall scattering, 
       they are considered lost. 
 \item \textbf{Emptying:} This is the counting phase, and it lasts for a certain time $t_E$. Only the UCN that in this phase hit the 
       trap wall in correspondence of the two valves that are connected with the two detectors (see fig. \ref{fig:5_trap}), 
       are considered counted and then taken into account for the calculation of $n_\ast$. 
       The exit valves have the same shape and dimension of entrance valve.
       UCN that decay, escape or do not hit the exit valves during $t_E$ are again considered lost.
\end{itemize}
We first computed the mean value of free flight time $t_{\rm f} = \langle t \rangle$, 
its variance $\langle t^2 \rangle$, etc. Their distribution were computed by averaging them over the individual UCN trajectories in the trap.
The parameters $v_{max}$ and $\eta$  in eq. (\ref{eq:escape}) were tuned to reproduce the experimental data such as the time constants for the neutron counts during 
the filling of the trap, UCN storage and the trap emptying. 
Our values are in good agreement with the parameters used in the previous experiments \cite{Serebrov,Serebrov-New} which used the same trap. 

Now we have all the instruments to compute the averaged number of collision $n_\ast$ for each experimental configuration.
We ran a thousand simulations per each experimental configuration, with $10^{5}$ neutron each, using the geometrical configuration of the trap 
of the experiments \cite{Serebrov,Serebrov-New,Nostro} and computed the averaged number of wall collisions with the criteria stated above.
Results are listed in table \ref{tab:nhit}.

We estimated the  uncertainties in the parameters $v_{max}$ and $\eta$ of eq. (\ref{eq:escape}) to be 10\% \cite{Ignatovich}. 
Thus for each simulation we randomly generated their values from a Gaussian distribution which have the parameters
used in \cite{Serebrov,Serebrov-New}, $\bar{v}_{max}= 6.8 m/s$ and $\bar{\eta}=2.2 \times 10^{-4}$, 
as mean values and standard deviations given by: $\sigma_v = \bar{v}_{max}/10$  and $\sigma_\eta = \bar{\eta}/10$ respectively.

\begin{table}[!ht]
\begin{center}
\begin{tabular}{cc}
\hline 
   ( Fill, Stor, Emt) [s] &   $n_*$ \\
 \hline 
 ( $100$, $250$, $150$)  & 2075 $\pm$ 86 \\
\hline 
 ( $130$, $300$, $130$)  & 2485 $\pm$ 103 \\
\hline
 ( $100$, $150$, $240$)  & 1489 $\pm$ 58 \\
\hline
\end{tabular}
\caption{Results of Monte Carlo simulation in different time duration of the experimental phases: 
filling(Fill), storage(Stor) and emptying(Emt).
The last line has been calculated using only one detector as in setup B4 of \cite{Nostro}.} 
\label{tab:nhit}
\end{center}
\end{table}

These numbers have been used to carry out the analysis in \cite{Nostro}. 
The configuration of the second line in table \ref{tab:nhit} is the same as in the experiment described in \cite{Serebrov-New}.
In their simulation, reported also in \cite{Berezhiani:2012rq} they found $n_\ast = 4000$. 

The discrepancy between our and their results can
be explained by the fact that we directly computed the number of collisions per each neutron considering only neutrons that have actually
been counted, while they calculated the UCN mean flight time $\langle t_f \rangle$ 
also including the not surviving neutrons and from that estimated $n_\ast$.

\section{Oscillation Probability}
With an in-homogeneous experimental magnetic field, to set limits on the oscillation time,
we cannot use the empirical formula used in \cite{Berezhiani:2012rq}, that is valid only in case of a uniform magnetic field.
We can use our simulation to estimate the dependence on the mirror magnetic field of the averaged value of the oscillation 
probability in the collision between UCN and the walls, in a specific experimental magnetic field configuration.
For using this in the analysis of \cite{Nostro}, we assumed that ordinary and mirror magnetic field are oriented in vertical direction.
To this aim, during the motion of a UCN between two consecutive collisions, the Schroedinger equation eq. (\ref{eq:schr}) has been numerically 
integrated.

The initial condition is set to have a neutron: $\Psi_0= (1 , 0)$, (pure neutron state), then, the wave function is 
computed in every temporal step using eq. (\ref{eq:Prob}).
When the UCN hit the wall, the program computes and saves the current value of the oscillation probability $P = \vert \psi_{n'} \vert^2$ 
and set $\Psi = \Psi_0$ back to the initial condition.

\begin{align}
  \psi_n(t_{i+1}) = & \cos\left( \delta \bar{\omega}_{i+1,i}   \; dt \right) \psi_n(t_i)  \nonumber \\
                    &-i\sin \left( \delta \bar{\omega}_{i+1,i} \; dt \right) \cos 2\theta_{i+1,i} \; \psi_n(t_i)  \nonumber \\
                    &-i\sin \left( \delta \bar{\omega}_{i+1,i} \; dt \right) \sin 2\theta_{i+1,i} \; \psi_{n'}(t_i)  \nonumber \\
\psi_{n'}(t_{i+1}) = & \cos\left( \delta \bar{\omega}_{i+1,i}   \; dt \right) \psi_{n'}(t_i)  \nonumber \\
                    &+i\sin \left( \delta \bar{\omega}_{i+1,i} \; dt \right) \cos 2\theta_{i+1,i} \; \psi_{n'}(t_i)  \nonumber \\
                    &-i\sin \left( \delta \bar{\omega}_{i+1,i} \; dt \right) \sin 2\theta_{i+1,i} \; \psi_n(t_i)   \label{eq:Prob} 
\end{align}
where 
\begin{align}
 \delta \bar{\omega}_{i+1,i} = & \; \sqrt{\delta\omega^2_{i+1,i} \, + \, \epsilon^2} \nonumber \\
        \delta\omega_{i+1,i} = & \; (\omega_{i+1} + \omega_i)/2 - \omega'  \nonumber \\
        \sin 2\theta_{i+1,i} = & \; \epsilon / \delta \bar{\omega}_{i+1,i}  \nonumber \\
        \cos 2\theta_{i+1,i} = & \; \delta\omega_{i+1,i} / \delta \bar{\omega}_{i+1,i} 
\end{align}
and the mirror magnetic field $\omega'$ (converted to the Larmor frequency) is assumed to be constant in the whole trap.
On the other hand $\omega_i$ is the value of the ordinary magnetic field in the position of the UCN at time $t_i$.
The magnetic field $\mathbf{B}$ was numerically calculated in every node of cubic lattice with 1 cm$^3$ elementary volumes, 
and in this way the distributions shown in figs. \ref{fig:dist} were obtained. 
In each elementary volume the magnetic field was taken as constant, with a value obtained by averaging between 
the values calculated at its vertices.

\begin{figure}[!ht]
 \begin{center} 
 \includegraphics[width=1.0\columnwidth]{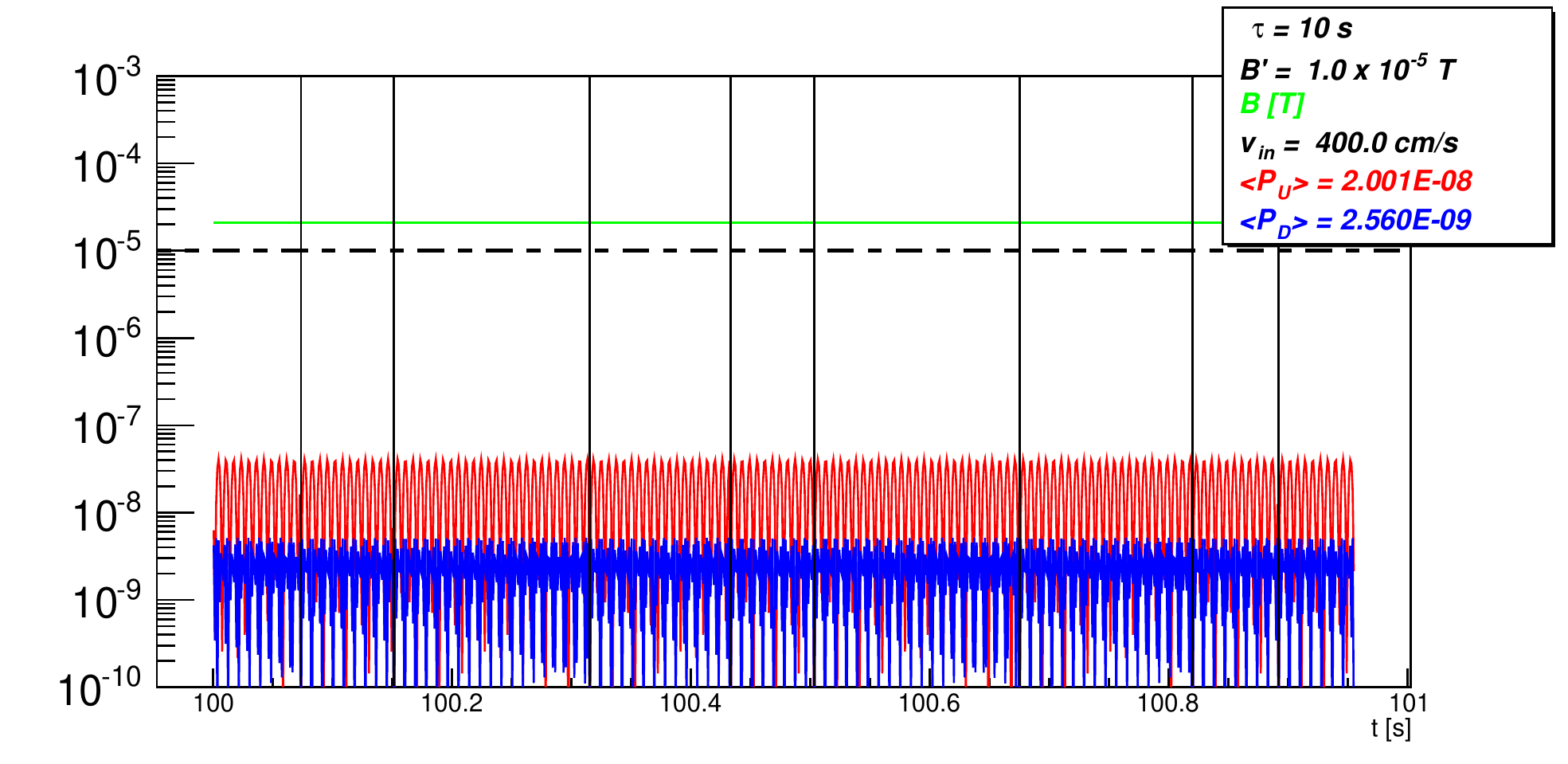} 
 \includegraphics[width=1.0\columnwidth]{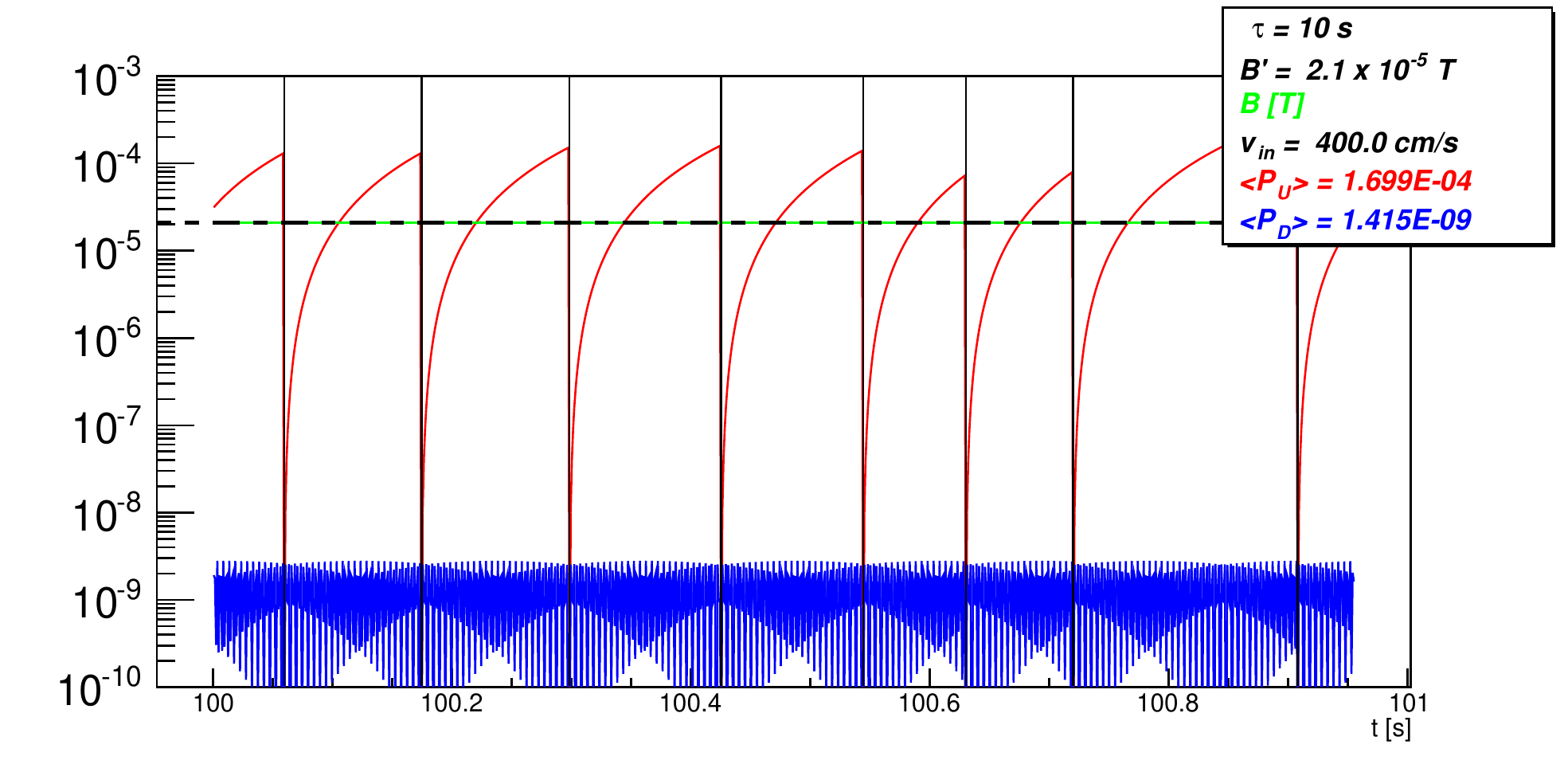}
 \includegraphics[width=1.0\columnwidth]{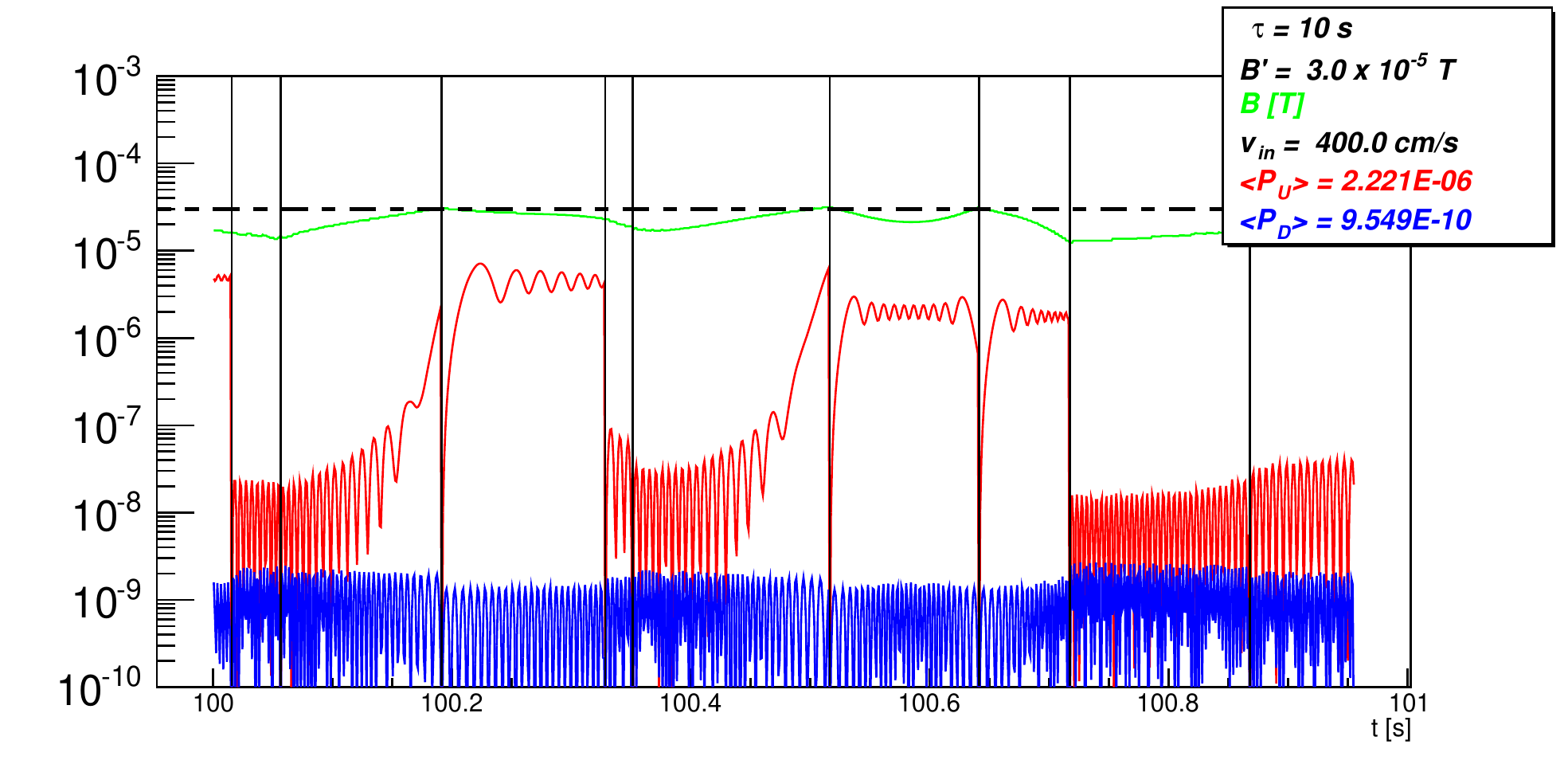}
 \includegraphics[width=1.0\columnwidth]{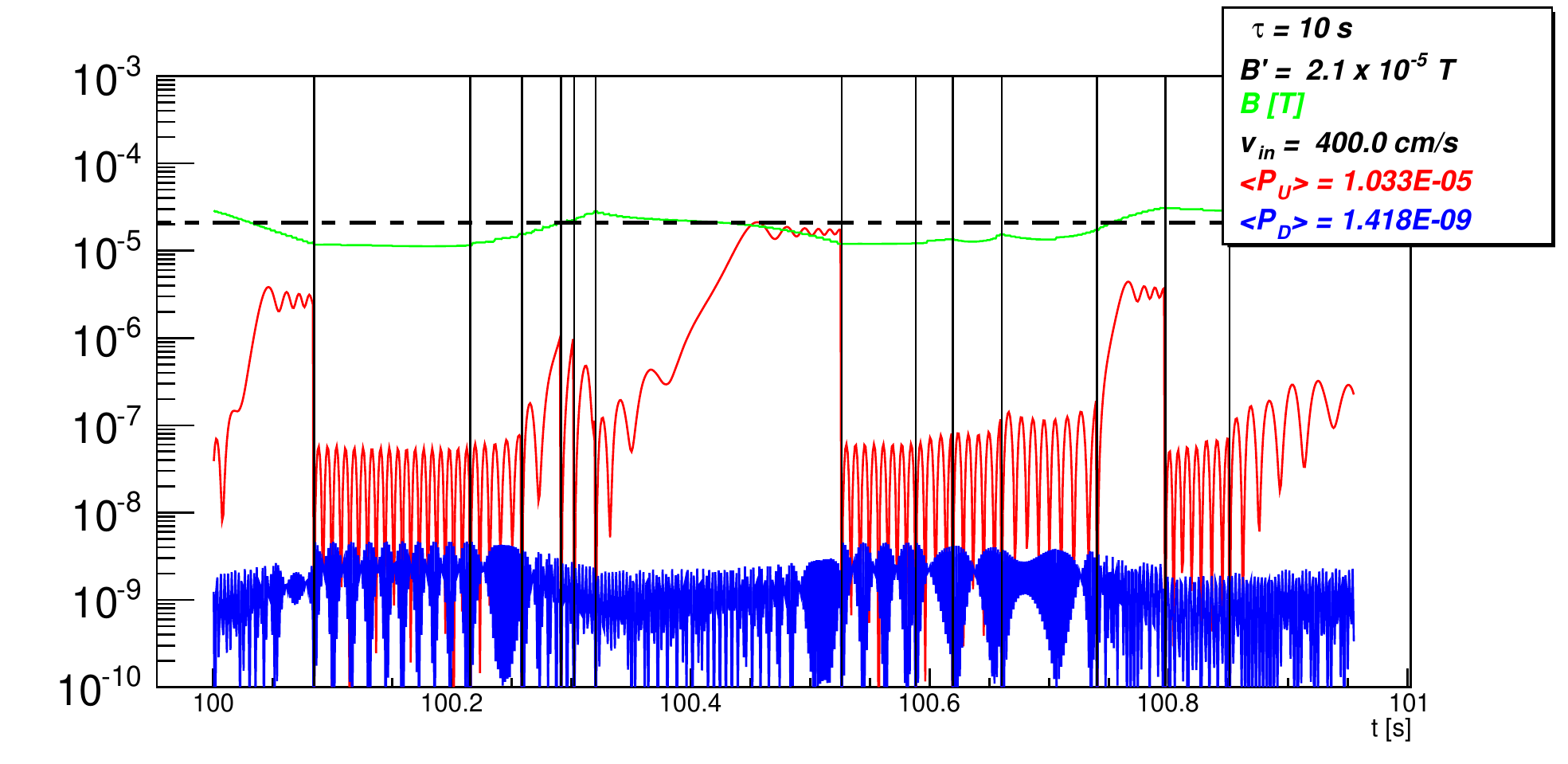} 
\caption{\label{fig:probability} Examples of behavior of $P_U$ (red line) and $P_D$ (blue line) in four different 
ordinary and mirror magnetic field 
configurations for a UCN with 4 m/s
velocity, assuming $\tau = 10$ s.
Black dashed line indicates $B'$ while green line indicates B in the region in which UCN is moving.
Each wall scattering is marked by a vertical black line.
  \textit{First panel:} uniform ordinary magnetic field (B = 0.21 G) far from the resonance ($B'$ = 0.1 G). 
  \textit{Second panel:} uniform ordinary magnetic at  resonance ($B'$ = B = 0.21 G)
  \textit{Third panel:} not uniform ordinary magnetic field far from resonance ($B_c$ = 0.21  G and $B'$ = 0.3 G).
  \textit{Fourth panel:} not uniform ordinary magnetic field close to resonance ($B_c$ = 0.21  G and $B'$ = 0.21 G).}
\end{center}
\end{figure}

Every UCN leaving an elementary cube with a given value of magnetic field with a wave vector $(\psi_n,\psi_{n'})$,  
while crossing the adjacent cube was evolving with the corresponding magnetic field.
In fig. \ref{fig:probability} we show some examples of the results of this calculation.
We computed simultaneously, in every temporal step, the values of the probabilities $P_U$ and $P_D$ which represent respectively the cases 
where the ordinary and mirror magnetic field are parallel ($\beta=0$) or antiparallel ($\beta=\pi$). 
As it can be easily understood from fig. \ref{fig:probability} as the UCN approaches to a region where the ordinary and mirror magnetic field
are in resonance (same direction and magnitude) 
the probability grows up to $10^{-4}$ in the case of ideal resonance, with $\tau = 10$ s.

\begin{figure}[!ht]
 \begin{center}
\includegraphics[width=1.\columnwidth]{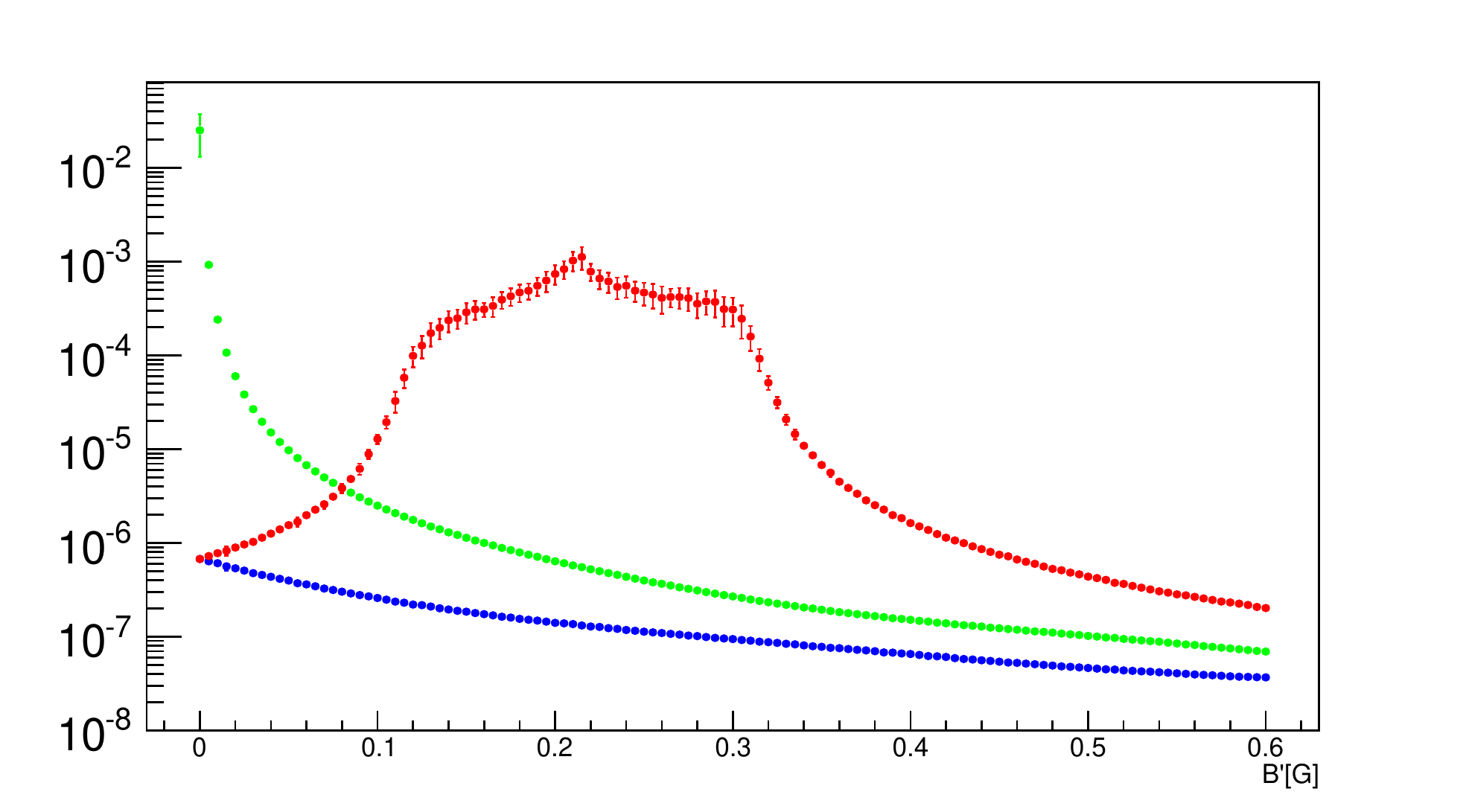}
\includegraphics[width=1.\columnwidth]{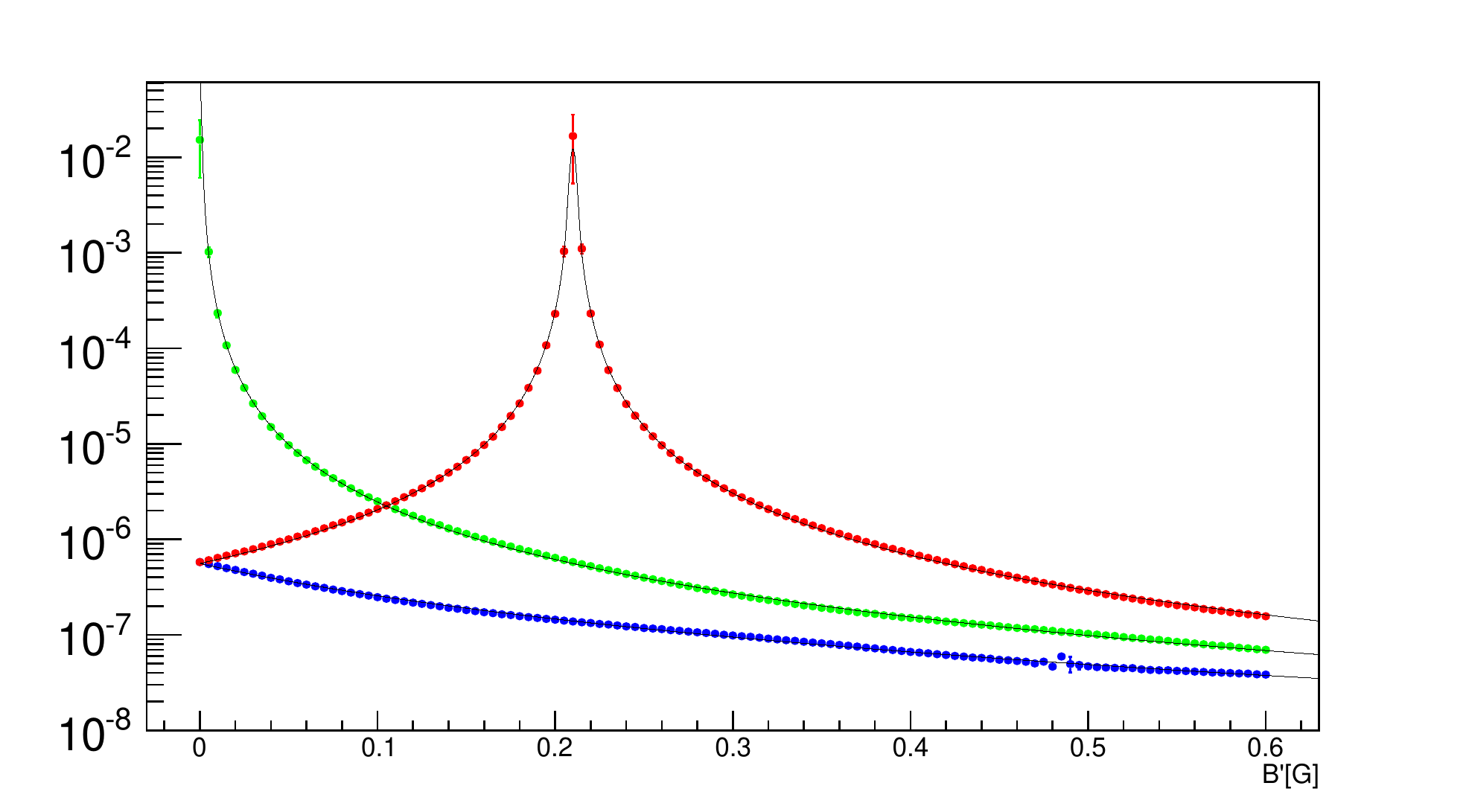}
\caption{\label{fig:simulation} Results of the simulation with vertical not uniform magnetic field ($B_c$ = 0.21 G).
\textit{Upper panel:}
in Red and blue we have, respectively, oscillation probability calculated in the case where B and $B'$ are parallel ($\beta=0$) or 
antiparallel ($\beta=\pi$) \textit{i.e.} $\mathcal{S}_{\pm}(B')$. 
In green the case where B = 0 ($S_0(B')$), error bars are given by the variance of the distributions (1 $\sigma$ ) for a given value of $B'$. 
\textit{Lower panel:}results of the simulation in the case of uniform experimental magnetic field (B = 0.21 G), compared with the
e approximation used in \cite{Berezhiani:2012rq} (Black lines). }
\end{center}
\end{figure}

Now we are able to study the dependence of the oscillation probabilities on the mirror magnetic field for a given profile of ordinary 
magnetic field. 
The UCN motion is integrated for 300 seconds in ideal conditions \textit{i. e.} no $\beta$-decay and no trap losses to maximize
hit statistics.
Since we are interested in oscillation probabilities for only the UCN that are going to be detected, the UCN initial spectra 
in this simulation has been modified taking into account the fact that neutron with higher velocity escape
from the trap after a small number of collisions due to eq. (\ref{eq:escape}).

We computed mean values $P_U(B')$ and $P_D(B')$ of the oscillation probabilities eq. (\ref{maxmin}) between wall scatterings, 
averaged over distribution of the neutron flight time $t$ and distribution of the magnetic field $B$ in the trap for a given value of $B_c$,
as functions of mirror magnetic field $B'$. 
In addition, we computed also the mean oscillation probability $P_0(B')$ for the case when no magnetic field was applied, $B=0$: 

\begin{align}
 \bar{P}_{U/D}(B') & =
\left\langle \frac{\sin^2[(\omega \mp \omega')t]}{\tau^2 (\omega \mp \omega')^2} \right\rangle_{t,B} = 
\left(\frac{1\, {\rm s}}{\tau}\right)^2 \! \mathcal{S}_{\pm}(B')  \, ,  \nonumber \\
 \bar{P}_0(B') & =  \left\langle \frac{\sin^2(\omega')t }{\tau^2 \omega^{\prime2}} \right\rangle_t
= \left(\frac{1\, {\rm s}}{\tau}\right)^2 \! \mathcal{S}_{0}(B') \, . 
\end{align}

In upper panel of fig. \ref{fig:simulation} we show simulation results for $\mathcal{S}_{\pm}(B')$ and $\mathcal{S}_0(B')$. 
These functions correspond to 
mean values of the respective probabilities in the case of $\tau=1$ s. 
For checking the consistency of our simulation, we also computed average oscillation 
probabilities in the case of homogeneous magnetic field in the same trap (lower panel of Fig. \ref{fig:simulation}). 
As we see, the results of our simulation, coincide with the results 
obtained via the empirical formula eq. (\ref{eq:empirical}) for the corresponding values of the parameters 
 $t_{\rm f}  = \langle t \rangle$ and $\sigma^2_{\rm f} = \langle t^2 \rangle - t_{\rm f}^2$, computed via our simulation.

From Fig. \ref{fig:simulation} we see that the in-homogeneous profile of the experimental magnetic fields has 
certain advantages: the function $\mathcal{S}_{+}(B') $ in homogeneous magnetic field has maximal sensitivity at the resonance, 
$B\approx B'$ as in the in-homogeneous case, but it is a very narrow resonance. 
So, if the variation of the field $B$ in search for the resonance is done with large steps, there is a chance to miss it.
On the other hand, in the in-homogeneous field setup, the enhancement of the oscillation probability covers much wider range of 
values of $B'$, and so, the search of the resonance can be done with larger steps, saving experimental time.

\section{Conclusions}
The aim of this work was to develop a tool that could be used in the analysis of any $n-n'$ oscillation in a UCN trap experiment using a 
non-uniform magnetic field and for comparing results from different $n-n'$ UCN oscillation experiments.
The code has been used in the analysis presented in \cite{Nostro}. 
The algorithm has been tested to maintain a good approximation up to 500 s UCN storage time with an energy loss less then 1 \%.
It has been developed for a cylindric trap, like the one used in \cite{Serebrov,Serebrov-New,Nostro}, but it can be easily modified for a 
different geometrical configuration such us spheroidal traps and
it can also simulate many different reflecting surfaces inside the trap. 
Unlike in the simulation used for \cite{Serebrov,Serebrov-New}, the calculation of the averaged number of collision $n_\ast$ have
not taken into account all the UCN entering the trap, but only the ones that do not decay or escape until they have been counted by the detectors.
We verified that the approximation for the calculation of oscillation probability used in \cite{Berezhiani:2012rq} is well reproduced by our 
simulation assuming an uniform ordinary magnetic field inside the trap.
The calculation of the $P_U$ and $P_D$ profiles show that experiments with not uniform magnetic field are sensitive to 
resonance in a larger set of $B'$ values. So, this kind of configuration can be used in future experiments to scan a larger set of values
for $B'$ looking for resonance.
Using this tools we could be able to compare UCN trapping experiment realized with different magnetic field, geometrical or technical setup. 
The code is available by contacting the author via email. 

\section{Acknowledgements} 
I want to thank Zurab Berezhiani and Nicola Rossi for useful discussions and technical support.


\begin{thebibliography}{99}

\bibitem{LY1} 
%
  I.~Y.~Kobzarev, L.~B.~Okun and I.~Y.~Pomeranchuk,
  Sov.\ J.\ Nucl.\ Phys.\  {\bf 3}, 
  837 (1966)
  [Yad.\ Fiz.\  {\bf 3}, 1154 (1966)];
\bibitem{LY2}
  R.~Foot, H.~Lew and R.~R.~Volkas,
  Phys.\ Lett.\ B {\bf 272}, 67 (1991).
 %
 
 
\bibitem{IJMPA1} 
  Z.~Berezhiani,
  Int.\ J.\ Mod.\ Phys.\ A {\bf 19}, 3775 (2004)
  [hep-ph/0312335]; 
\bibitem{IJMPA2} 
  Z.~Berezhiani,
  Eur.\ Phys.\ J.\ ST {\bf 163}, 271 (2008); 
\bibitem{IJMPA3} 
  Z.~Berezhiani,
  In Shifman, M. et al. (eds.): {\it From fields to strings}, vol. 3, pp. 2147-2195, 
  doi:10.1142/9789812775344-0055
  [hep-ph/0508233]; 
\bibitem{IJMPA4} 
  R.~Foot,
  Int.\ J.\ Mod.\ Phys.\ A {\bf 29}, 1430013 (2014)
  [arXiv:1401.3965 [astro-ph.CO]].
 

\bibitem{BDM1} 
  Z.~Berezhiani, A.~D.~Dolgov and R.~N.~Mohapatra,
  Phys.\ Lett.\ B {\bf 375}, 26 (1996); 
\bibitem{BDM2}
  Z.~G.~Berezhiani,
  Acta Phys.\ Polon.\ B {\bf 27}, 1503 (1996)
  [hep-ph/9602326].
%

\bibitem{BCV1} 
  Z.~Berezhiani, D.~Comelli and F.~L.~Villante,
  Phys.\ Lett.\ B {\bf 503}, 362 (2001)
  [hep-ph/0008105]; 
\bibitem{BCV2} 
  A.~Y.~Ignatiev and R.~R.~Volkas,
  Phys.\ Rev.\ D {\bf 68}, 023518 (2003)
  [hep-ph/0304260]; 
\bibitem{BCV3} 
  Z.~Berezhiani, P.~Ciarcelluti, D.~Comelli and F.~L.~Villante,
  Int.\ J.\ Mod.\ Phys.\ D {\bf 14}, 107 (2005)
  [astro-ph/0312605]; 
\bibitem{BCV4} 
  Z.~Berezhiani, S.~Cassisi, P.~Ciarcelluti and A.~Pietrinferni,
  Astropart.\ Phys.\  {\bf 24}, 495 (2006)
  [astro-ph/0507153].

 
\bibitem{PLB981} 
  B.~Holdom,
  Phys.\ Lett.\  {\bf 166B}, 196 (1986); 
\bibitem{PLB982} 
  Z.~Berezhiani,
  Phys.\ Lett.\ B {\bf 417}, 287 (1998);
\bibitem{PLB983} 
  Z.~Berezhiani, L.~Gianfagna and M.~Giannotti,
  Phys.\ Lett.\ B {\bf 500}, 286 (2001)
  [hep-ph/0009290].
 \bibitem{PLB984} 
  A.~Addazi, Z.~Berezhiani and Y.~Kamyshkov,
  Eur.\ Phys.\ J.\ C {\bf 77}, no. 5, 301 (2017)
  [arXiv:1607.00348 [hep-ph]]; 
 \bibitem{PLB985} 
  Z.~Berezhiani,
  Eur.\ Phys.\ J.\ C {\bf 76}, no. 12, 705 (2016)
  [arXiv:1507.05478 [hep-ph]].
 
 

\bibitem{DAMA1} 
  R.~Foot,
  Int.\ J.\ Mod.\ Phys.\ A {\bf 19}, 3807 (2004)
  [astro-ph/0309330]; 
 \bibitem{DAMA2} 
  Phys.\ Rev.\ D {\bf 86}, 023524 (2012)
  [arXiv:1203.2387 [hep-ph]]; 
 \bibitem{DAMA3} 
  R.~Cerulli, {\it et al.}, 
  Eur.\ Phys.\ J.\ C {\bf 77}, no. 2, 83 (2017)
  [arXiv:1701.08590 [hep-ex]]; 
\bibitem{DAMA4} 
  A.~Addazi, {\it et al.}, 
  Eur.\ Phys.\ J.\ C {\bf 75}, no. 8, 400 (2015)
  [arXiv:1507.04317 [hep-ex]].
 
\bibitem{M-neutrinos1} 
  R.~Foot, H.~Lew and R.~R.~Volkas,
  Mod.\ Phys.\ Lett.\ A {\bf 7}, 2567 (1992); 
\bibitem{M-neutrinos2} 
  E.~K.~Akhmedov, Z.~G.~Berezhiani and G.~Senjanovic,
  Phys.\ Rev.\ Lett.\  {\bf 69}, 3013 (1992); 
\bibitem{M-neutrinos3} 
 R.~Foot and R.~R.~Volkas,
  Phys.\ Rev.\ D {\bf 52}, 6595 (1995); 
 \bibitem{M-neutrinos4} 
  Z.~Berezhiani and R.~N.~Mohapatra,
  Phys.\ Rev.\ D {\bf 52}, 6607 (1995);  



\bibitem{BB-PRL1} 
  L.~Bento and Z.~Berezhiani,
  Phys.\ Rev.\ Lett.\  {\bf 87}, 231304 (2001)
  [hep-ph/0107281]; 
 \bibitem{BB-PRL2} 
  L.~Bento and Z.~Berezhiani,
  Fortsch.\ Phys.\  {\bf 50}, 489 (2002); 
 \bibitem{BB-PRL3} 
  L.~Bento and Z.~Berezhiani,
  hep-ph/0111116; 
\bibitem{BB-PRL4} 
  Z.~Berezhiani,
  arXiv:1602.08599 [astro-ph.CO].
  
 


\bibitem{BB-nn'} 
  Z.~Berezhiani and L.~Bento,
  Phys.\ Rev.\ Lett.\  {\bf 96}, 081801 (2006)
  [hep-ph/0507031].  
  
\bibitem{More} 
  Z.~Berezhiani,
  Eur.\ Phys.\ J.\ C {\bf 64}, 421 (2009)
  [arXiv:0804.2088 [hep-ph]].
 

\bibitem{Phillips1}  
  D.~G.~Phillips {\it et al.},
  Phys.\ Rept.\  {\bf 612}, 1 (2016)
  [arXiv:1410.1100 [hep-ex]];
 \bibitem{Phillips2} 
  K.~S.~Babu {\it et al.},
  ``Neutron-Antineutron Oscillations: A Snowmass 2013 White Paper,''
  arXiv:1310.8593 [hep-ex];
 \bibitem{Phillips3} 
  M. Baldo-Ceolin {\it et al.}, 
  Z. Phys. C {\bf 63}, 409 (1994).

  
\bibitem{UHECR1} 
  Z.~Berezhiani and L.~Bento,
  Phys.\ Lett.\ B {\bf 635}, 253 (2006)
  [hep-ph/0602227]; 
\bibitem{UHECR2} 
  Z.~Berezhiani and A.~Gazizov,
  Eur.\ Phys.\ J.\ C {\bf 72}, 2111 (2012)
  [arXiv:1109.3725 [astro-ph.HE]]; 
  

\bibitem{Pokot} 
  Y.~N.~Pokotilovski,
  Phys.\ Lett.\ B {\bf 639}, 214 (2006)
  [nucl-ex/0601017].
  
\bibitem{ORNL1}  
  Z.~Berezhiani {\it et al.}
  Phys.\ Rev.\ D {\bf 96}, no. 3, 035039 (2017)
  [arXiv:1703.06735 [hep-ex]]; 
\bibitem{ORNL2} 
  L.~J.~Broussard {\it et al.},
  arXiv:1710.00767 [hep-ex].
  

\bibitem{Bison} 
  G.~Bison {\it et al.},
  Phys.\ Rev.\ C {\bf 95}, no. 4, 045503 (2017)
  [arXiv:1610.08399 [physics.ins-det]].
 

\bibitem{Ban} 
  G.~Ban {\it et al.},
  Phys.\ Rev.\ Lett.\  {\bf 99}, 161603 (2007)
  [arXiv:0705.2336 [nucl-ex]].
  
\bibitem{Serebrov} 
  A.~P.~Serebrov {\it et al.},
  Phys.\ Lett.\ B {\bf 663}, 181 (2008)
  [arXiv:0706.3600 [nucl-ex]].
  
\bibitem{Serebrov-New} 
  A.~P.~Serebrov {\it et al.},
  Nucl.\ Instrum.\ Meth.\ A {\bf 611}, 137 (2009)
  [arXiv:0809.4902 [nucl-ex]].
  
  
 \bibitem{Bodek} 
 K.~Bodek {\it et al.},
  Nucl.\ Instrum.\ Meth. \ A {\bf 611}, 141 (2009).

  
\bibitem{Altarev} 
  I.~Altarev {\it et al.},
  Phys.\ Rev.\ D {\bf 80}, 032003 (2009)
  [arXiv:0905.4208 [nucl-ex]].
  
\bibitem{Nostro} 
  Z.~Berezhiani {\it et al.} 
  arXiv:1712.05761 [hep-ex].
  
  
\bibitem{PDG} 
  C.~Patrignani {\it et al.} [Particle Data Group],
  Chin.\ Phys.\ C {\bf 40}, no. 10, 100001 (2016).
    
  

\bibitem{Berezhiani:2012rq} 
  Z.~Berezhiani and F.~Nesti,
  Eur.\ Phys.\ J.\ C {\bf 72}, 1974 (2012)
  [arXiv:1203.1035 [hep-ph]].
  

  
\bibitem{BDT}  
  Z.~Berezhiani, A.~D.~Dolgov and I.~I.~Tkachev,
  Eur.\ Phys.\ J.\ C {\bf 73}, 2620 (2013)
  [arXiv:1307.6953 [astro-ph.CO]].
 
 
  
\bibitem{UCN_spectra}
 A. Steyerl et al.,  Phys. Lett. A {\bf 116}, 347 (1986).

\bibitem{UCN_reflection}
 F. Atchison et al.,
 Nuclear Instruments and Methods in Physics Research A 552 (2005) 513–52
  
\bibitem{Golub}
  R.~Golub and J.~M.~Pendlebury,
  Rept.\ Prog.\ Phys.\  {\bf 42}, 439 (1979).
  
  
\bibitem{Ignatovich}
V.K. Ignatovich, The Physics of Ultracold Neutrons, Clarendon Press (1990), Oxford, UK



\end{thebibliography}
\end{document}